\documentclass[twocolumn,amsmath,amssymb,
showpacs,
 aps,
prb,
 lengthcheck,%
]{revtex4-1}
\usepackage{graphicx}
\usepackage{dcolumn}
\usepackage{bm}
\usepackage{hyperref}
\usepackage{makeidx}
\usepackage{ulem}

\usepackage[english]{babel} 
\usepackage[utf8]{inputenc}
\usepackage{color}
\usepackage{graphicx}
\usepackage{latexsym}
\usepackage{amsthm}
\usepackage{dsfont}

\definecolor{mygreen}{rgb}{0.0,0.5,0.0}

\newcommand{\cG}{{\cal G}}

\makeindex

\begin{document}
 
\title{Finite frequency noise in a normal metal - topological superconductor junction}

\author{D. Bathellier,$^1$ L. Raymond,$^1$ T. Jonckheere,$^1$ J. Rech,$^1$ A. Zazunov,$^2$ and T. Martin$^1$}
\affiliation{${}^1$~Aix Marseille Univ, Universit\'e de Toulon, CNRS, CPT, Marseille, France}
\affiliation{${}^2$~Institut f\"ur Theoretische Physik, Heinrich Heine Universit\"at, D-40225 D\"usseldorf, Germany}

\date{\today}

\begin{abstract}
A topological superconductor nanowire bears a Majorana bound state at each of its ends, leading to unique transport properties. As a way to probe these, we study the finite frequency noise of a biased junction between a normal metal and a topological superconductor nanowire.  We use the non-equilibrium Keldysh formalism  to compute the finite frequency emission and absorption noise to all order in the tunneling amplitude, for bias voltages below and above the superconducting gap. We observe noticeable structures in the absorption and emission noise, which we can relate to simple transport processes. The presence of the Majorana bound state is directly related to a characteristic behavior of the noise spectrum at low frequency. We further compute the noise measurable with a realistic setup, based on the inductive coupling to a resonant LC circuit, and discuss the impact of the detector temperature. We have also computed the emission noise for a non-topological system with a resonant level, exhibiting a zero-energy Andreev bound state, in order to show the specificities of the topological case. Our results offer an original  tool for the further characterization of the presence of Majorana bound states in condensed matter systems. 
\end{abstract}

\maketitle

\section{Introduction}

The search for Majorana fermions \cite{Majorana_37} -- particles that are their own antiparticles -- has been an active field of study in high energy physics, with no conclusive/definite evidence of their finding. Since the last decades, the Majorana paradigm has now entered condensed matter physics thanks to the pioneering work of Kitaev. \cite{kitaev_01} In this context, rather than being elementary particles, Majorana fermions emerge from the collective behavior of a many electron system. This Kitaev toy model consists in a tight-binding chain of electrons living on lattice sites, which includes hopping and superconducting (p-wave) pairing between neighboring sites. Majorana fermions exist at the boundaries of this one dimensional chain provided that certain conditions on the hopping parameter, the gap parameter, and the chemical potential are met. In this so-called topological phase, the ground state of the system is degenerate, opening the possibility to realize a quantum bit which could be in principle immune from decoherence effects, with applications to quantum information schemes. 

In quantum nano-physics, a huge effort has been devoted to finding a realistic experimental equivalent of the Kitaev model. One possibility is to realize a topological superconductor (TS) nanowire by inducing s-wave superconductivity in a semiconducting nanowire subject to both Rashba spin-orbit coupling and a Zeeman magnetic field. \cite{lutchyn_10,oreg_01,alicea_11,mourik_12,das_12,lutchyn18} Other proposals where a chain of magnetic atoms are deposited on a superconducting substrate have also been studied.\cite{nadj-perge_14} Many theoretical proposals for the detection of Majorana fermions relying on quantum transport setups have been published over the past few years, \cite{alicea_12,leijnse_12,beenakker_13,stanescu_13,aguado_17,elliott_15} but despite the significant experimental progress, an unequivocal signature\cite{kashiwaya_00,burset_17} of these Majorana bound states (MBS) is still lacking. While some elements point in the right direction (the detection of the zero bias anomaly in the differential conductance, and quite recently a claim that the quantized zero bias conductance has been observed \cite{zhang_18}) it is clear that the definite discovery of Majorana fermions will require more experimental tests in order for this finding to be firmly established. This is especially important in order to exploit the properties of these objects in the future, such as the generation of a Majorana qbit, and the braiding of Majorana fermions, with applications to quantum computing.

\begin{figure}[tb]
\centering
\includegraphics[width=\columnwidth]{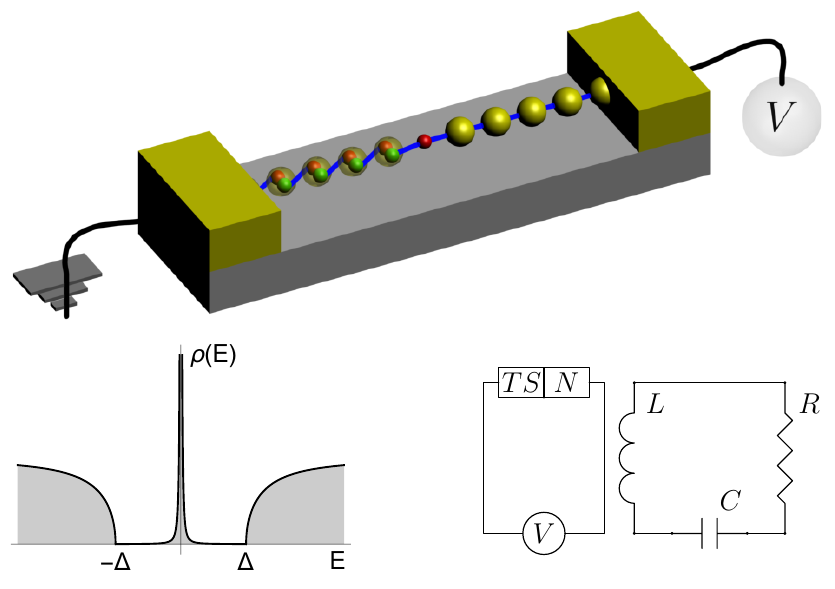}
\caption{Top:  schematic view of a junction between a normal metal (yellow dots) and a topological superconductor (red-green dots, with a red Majorana bound state at its end);
the normal metal is biased at voltage $V$. Bottom left: topological superconductor density of states, with a central peak at zero energy due to the Majorana bound state, and square root behavior outside the gap.
Bottom right: the principle of the noise detection, using the inductive coupling to a RLC circuit.\cite{lesovik_97}}
\label{fig:system}
\end{figure}

In this work, the physics of a normal metal - topological superconductor (NTS) junction (see Fig.~\ref{fig:system}) is revisited. While the non-equilibrium electronic current and the zero-frequency noise are perfectly understood by now, \cite{zazunov_16} the finite frequency noise characteristic of a NTS junction has so far not been considered. Finite frequency noise has been studied extensively in more conventional materials in the context of quantum mesoscopic physics and nano-physics, both theoretically and experimentally. On the theoretical side, it has been argued\cite{lesovik_97,gavish_00,aguado_00} that since the two current operators entering the noise correlator are evaluated at different times, one can introduce two quantities - the emission and the absorption noise. A noise detector should in principle be described on the same quantum footing as the nano-device under study. In full generality, the Fourier transform of the real time noise correlators can be a complex function of the frequency. The measured signal by the quantum detector, which is a real number, should thus be a combination of such emission and absorption quantities. This has been experimentally verified in pioneering finite frequency measurement experiments performed in Deflt,\cite{deblock_03} in Orsay\cite{billangeon_06,billangeon_07,basset_10} and in Saclay\cite{zakka-bajjani_07,hofheinz_11,altimiras_14} in different contexts. In a previous work, one of us \cite{lesovik_99,torres_01} computed the finite frequency noise of a normal metal - BCS superconductor junction, and found cusps of singularities at specific frequencies corresponding to different transport processes. However, these results are not expected to survive in the case of the NTS junction, as the density of states (DOS) of the topological superconductor differ substantially from its conventional BCS counterpart. The BCS density of states is zero inside the gap, and bears inverse square root divergences at the gap edges. On the opposite, the DOS of a TS wire has a square root behavior outside the gap and a so-called Majorana peak at zero energy (see Fig.~\ref{fig:system}).    

Using a continuous model describing a semi-infinite Kitaev chain, we derive the finite frequency noise to all orders in the tunnel coupling parameter of the NTS junction (recovering in the process known results for the current and zero-frequency noise\cite{zazunov_16}). Within our framework, the current and noise can be expressed in terms of dressed Keldysh Green's functions (using Wick's theorem for the noise as the Hamiltonian is quadratic). Solving Dyson's equation exactly, one reaches analytical expressions for these two quantities in terms of the known bare boundary Green's functions of the two leads. The finite frequency emission and absorption noises are then computed for voltages above and below the superconducting gap, showing plateaus - with a specific frequency width - where the latter quantities are essentially constant. We interpret such features as emission and absorption processes which couple the constant density of states of a normal metal with the sharp peak in the TS density of states associated with the Majorana bound states. For completeness, we further use a generic model for the quantum detector - a LC circuit which can absorb or emit photons from or to the nano-device\cite{lesovik_97,gavish_00,zazunov_07} - and compute the measurable noise at the resonant frequency of the LC detector.  

In order to illustrate the added value of using finite frequency noise over other quantities relying on DOS, like the differential conductance, we have also computed the transport properties of a non-topological 
N-dot-S system, where S is a standard BCS superconductor. Indeed this system bears a zero-energy Andreev bound state, with a sharp peak in the DOS, which is susceptible to yield similar differential conductance and finite frequency noise profiles.
With the right parameters, this system presents a conductance peak similar to the one
of a Majorana bound state, which shows that the conductance alone is not enough to characterize
a MBS. However, the emission noise shows a distinctive dip near zero-frequency in this system, which is absent for the NTS junction.
We thus believe that the finite frequency noise characteristics of an NTS junction can bring unique, additional information about Majorana fermions in experiments. This constitutes the main motivation of this study.

The paper is organized as follows: Sec.~\ref{sec:model} introduces the microscopic model Hamiltonian and defines the Keldysh Green's functions. The Dyson's equations which are relevant for the current and noise calculations are specified in Sec.~\ref{sec:dyson}. The noise is computed in Sec.~\ref{sec:noise} (real time correlator, emission and absorption noise,...) and detailed results are discussed in Sec.~\ref{sec:numerical}. Sec.~\ref{sec:NdS} shows the calculations and the results for the emission noise of a N-dot-S system, with a BCS superconductor. We conclude this work with Sec.~\ref{sec:conclusion}. In the whole following text, we fix the value of $\hbar$ to $1$.

\section{Model}
\label{sec:model}

A semi-infinite topological superconductor wire is a single-channel p-wave superconductor with spin-momentum locking, described from the continuous version of the Kitaev chain\cite{kitaev_01} by the helical superconductor Hamiltonian
\begin{equation}
H_j = \displaystyle\int_{0}^{+\infty} \mathrm{d}x ~\psi_{j}^{\dagger}(x) (-i v_F \partial_x \sigma_3 + \Delta \sigma_2)\psi_{j}(x)~,
\end{equation}
where $v_F$ is the Fermi velocity, $\Delta$ is the superconducting gap,  and $\sigma_{1,2,3}$ are Pauli matrices in Nambu space. Here, we refer to $j=0$ for the TS nanowire and $j=1$ for the (spinless) normal metal lead, whose Hamiltonian has the same form, with the gap parameter set to zero. These lead Hamiltonians are conveniently written in terms of the fermionic Nambu spinor $\psi_j$ defined as
\begin{equation}
 \psi_j= \begin{pmatrix}
	c_{Rj}\\[2mm]
	c_{Lj}^{\dagger}\\[2mm]
\end{pmatrix}~,
\end{equation}
which combines left- and right moving fermion operators $c_{R/L,j}$.

The Hamiltonian describing the tunnel coupling between the two leads has the form (in the absence of applied bias) 
\begin{equation}
H_T = \lambda c_0^{\dagger} c_1 + \mathrm{H.c.} ~,
\label{tunnelHamiltonian}
\end{equation}
where $c_0$ and $c_1$ are boundary fermions built simply from the sum of the left and right fermionic fields entering the Nambu spinor, evaluated at the location $x=0$ of the NTS junction, so that $c_j = c_{Rj} (0) + c_{Lj} (0)$.  
It is also convenient to write this tunnel Hamiltonian in the Nambu representation. 
The formalism can be generalized to an arbitrary number $M$ of tunnel-coupled conductors as in Ref. \onlinecite{jonckheere_17}. For this, we define a tunneling matrix $W$ in lead space, whose diagonal elements vanish: $W_{jj'} = \lambda \sigma_3 \left( 1 - \delta_{jj'} \right)$. Then, in Nambu representation, the tunneling Hamiltonian takes the form 
\begin{equation}
H_T = \dfrac{1}{2} \displaystyle\sum_{jj'}^{M} \Psi_j^{\dagger} W_{jj'} \Psi_{j'} ~~,~~~~\Psi_j = 
\begin{pmatrix}
c_{j}\\[2mm]
c_{j}^{\dagger} \\[2mm]
\end{pmatrix}~.
\label{tunnelHamiltonian2}
\end{equation}

The operator accounting for the current flowing through lead $j$, is derived from the tunneling Hamiltonian \eqref{tunnelHamiltonian2} as
\begin{equation}
I_j = i e \displaystyle\sum_{j'\neq j} \Psi^{\dagger}_j \sigma_3 W_{jj'} \Psi_{j'}~.
\label{currentop}
\end{equation}

An essential tool for this study is the Keldysh Green's function for the boundary fermions. It is defined as $\tilde{G}^{\eta \eta'}_{jk}(t,t') = -i \left\langle T_K  c_j(t_{\eta}) c_k^{\dagger}(t_{\eta'}) \right\rangle$, where $\eta = \pm$ denotes the Keldysh time index which specifies the position of times on the Keldysh contour. It is represented by the (4 by 4) matrix in Keldysh-Nambu space: 
\begin{equation}
\tilde{G}_{jj'} = \begin{pmatrix}
  G_{jj'}^{++} &  G_{jj'}^{+-}   \\
  G_{jj'}^{-+} &   G_{jj'}^{--}   \\   
\end{pmatrix} = \tilde{L}   \begin{pmatrix}
  0 &  G_{jj'}^a  \\
  G_{jj'}^r &   G_{jj'}^K   \\   
\end{pmatrix} \tilde{L}^{-1}~,
\label{GF}
\end{equation}
where $G^a$, $G^r$, $G^K$ are the advanced, retarded and Keldysh components in Nambu space, and 
the unitary transform $\tilde{L} = (1/\sqrt{2})(\tau_0+i\tau_2) \otimes \sigma_0$ ($\tau_j$ are Pauli matrices in Keldysh space). 

We wish to express the average current in terms of such Green's functions. For this we write explicitly the expression of the off-diagonal Keldysh Green's functions:
\begin{equation}
G^{-+}_{jj'}(t,t') = -  i \begin{pmatrix}
\langle c_{j}(t) c_{j'}^{\dagger}(t')\rangle & \langle c_j (t) c_{j'}(t')\rangle \\
\langle c_{j}^{\dagger}(t) c_{j'}^{\dagger}(t')\rangle & \langle c_{j}^{\dagger}(t) c_{j'}(t')\rangle  \\
\end{pmatrix} 
\label{greenfunctionmp}
\end{equation}

\begin{equation} 
G^{+-}_{jj'}(t,t') = + i \begin{pmatrix}
\langle c_{j'}^{\dagger}(t') c_{j}(t)\rangle & \langle c_{j'} (t') c_{j}(t)\rangle \\
\langle c_{j'}^{\dagger}(t') c_{j}^{\dagger}(t)\rangle & \langle c_{j'}(t') c_{j}^{\dagger}(t)\rangle   \\
\end{pmatrix}\label{greenfunctionpm}
\end{equation}

The average current can then be recast\cite{zazunov_16} in the form of a Nambu trace containing the Keldysh Green's function $G^K_{jj'}(t,t')= G^{+-}_{jj'}(t,t')+G^{-+}_{jj'}(t,t')$: 
\begin{equation}
\langle I_j\rangle = \dfrac{e}{2} \displaystyle\sum_{j'\neq j} \mathrm{Tr_N} \left[ \sigma_3 W_{j j'} G^K_{j'j}(t,t) \right] ~,
\end{equation}
which reads, in Fourier space, for the two-lead geometry chosen here
\begin{equation}
I(V) = \dfrac{e \lambda}{2} \displaystyle\int_{-\infty}^{+\infty} \dfrac{\mathrm{d}\omega}{2\pi} \mathrm{Tr_N} G_{01}^{K}(\omega)~,
\label{tracecourant}
\end{equation}
where we used that, in this case, the tunneling matrix reduces to $W_{01} = W_{10} = \lambda \sigma_3$.


\section{Dyson's equations and average current}
\label{sec:dyson}

The Dyson's equation is readily expressed in Keldysh-Nambu-lead space as
\begin{equation}
\check{G} = \left[ \check{g}^{-1} - \check{W} \right]^{-1}~,
\end{equation}
where we have
\begin{equation}
\check{G} = 
\begin{pmatrix}
\tilde{G}_{00} & \tilde{G}_{01} \\
\tilde{G}_{10} & \tilde{G}_{11}  \\
\end{pmatrix} 
~~,~~~~
\check{W} = 
\begin{pmatrix}
\tilde{W}_{00} & \tilde{W}_{01} \\
\tilde{W}_{10} & \tilde{W}_{11}  \\
\end{pmatrix} 
~,
\end{equation}
and the tunneling self-energy $\tilde{W}_{jj'}$ is a diagonal matrix of the form $\tilde{W}_{jj'} = \text{diag} \left( W_{jj'}, -W_{jj'}\right)$.

Using Eq.~\eqref{GF}, this can be further expressed into a set of equations for the advanced, retarded and Keldysh Green's functions as
\begin{align}
\label{eq:Gra}
\hat{G}^{r/a} &= \hat{g}^{r/a} + \hat{g}^{r/a} \hat{W} \hat{G}^{r/a} \\
\hat{G}^K &= \left( \hat{\mathds{1}} + \hat{G}^r \hat{W} \right) \hat{g}^K \left( \hat{\mathds{1}} + \hat{W} \hat{G}^a \right)
\label{eq:GK}
\end{align}
where we used that the Keldysh component associated with the tunneling self-energy vanishes and introduced the notation $\hat{\cdot}$ for matrices in Nambu-lead space.

In order to compute the current, we thus need to know the full expression of the dressed Green's function $G_{01}^K$. The Dyson's equation allows us to express this Keldysh component in terms of the advanced and retarded dressed Green's functions as well as the bare Keldysh Green's function and the tunneling self-energy. One has
\begin{align}
G_{01}^K &= \left(\sigma_0 + G_{01}^r W_{10}\right)g_{00}^K W_{01} G^{a}_{11}\nonumber  \\
&+ G_{00}^r W_{01}g_{11}^K\left(\sigma_0 + W_{10} G_{01}^{a}\right)~.
\label{GK}
\end{align}

\subsection{Bare Green's functions}

As a first step, we need to specify $g^K_{00}$ and $g^K_{11}$. 
The procedure to obtain the boundary Green's function for the uncoupled semi-infinite TS wire has been described carefully in Ref. \onlinecite{zazunov_16}. Following this derivation, the various components of the bare Green's functions for the topological superconductor then read
\begin{align}
g_{00}^K(\omega) &= \left[1 - 2 n_F(\omega)\right]\left[g_{00}^r(\omega) - g_{00}^a(\omega)\right] \\
g_{00}^{r/a}(\omega) &= \dfrac{\sqrt{\Delta^2 - (\omega \pm i0^{+})^2} \sigma_0 + \Delta \sigma_1 }{\omega \pm i0^+}~,
\end{align}
where $n_F (\omega)$ is the Fermi function and $0^+$ an infinitesimal introduced to properly define the square root in the complex plane. 

The Keldysh, retarded and advanced bare Green's functions for the metallic lead are readily obtained by setting $\Delta=0$ in the results above, so that 
\begin{align}
g_{11}^{K}(\omega)  &=  -2 i F_1 \\
g_{11}^{r/a}(\omega) & = \mp i \sigma_0 \label{g11}~.
\end{align}
There we have introduced the matrix $F_1$ as
\begin{equation}
F_1  =  \begin{pmatrix}
1- 2n_F(\omega-eV) &  0 \\[2mm]
0 & 1-2n_F(\omega+eV) 
\end{pmatrix}~.
\end{equation}
Note that the top and bottom diagonal elements respectively have an electron-like and a hole-like character, thus leading to opposite signs for the voltage.

Introducing the Heaviside distribution $\Theta(|\omega|-\Delta)$, the boundary Keldysh Green's function on the TS side becomes
\begin{align}
g_{00}^K(\omega) &= 2 i  \tanh\left(\frac{\beta\omega}{2}\right) \nonumber\\
\times &\left[ \pi \Delta (\sigma_0 + \sigma_1) \delta(\omega) + \dfrac{\sqrt{\omega^2 - \Delta^2}}{|\omega|}  \Theta(|\omega|-\Delta) \sigma_0\right]~.
\end{align}
where $\beta$ is the inverse temperature.

\subsection{Average current}

In order to write the average current, we then need the expression for the dressed Green's functions $G_{00}^{r/a}$, $G_{11}^a$ and $G_{00}^r$. 
Using Dyson's equation for the full advanced and retarded Green's function, Eq.~\eqref{eq:Gra}, it is convenient to express $G_{01}^{r/a}$, $G_{10}^{r/a}$ and $G_{11}^{r/a}$ in terms of $G_{00}^{r/a}$ as
\begin{align}
G_{01}^{r/a}(\omega) & = \mp i \lambda G_{00}^{r/a}(\omega) \sigma_3 \label{G01}~,\\
G_{10}^{r/a}(\omega) & = \mp i \lambda  \sigma_3 G_{00}^{r/a}(\omega) \label{G10}~,\\
G_{11}^{r/a}(\omega) & =\mp i \sigma_0 -\lambda^2 \sigma_3 G_{00}^{r/a}(\omega) \sigma_3~.  \label{G11}
\end{align}
It follows that the Keldysh component of the dressed Green's function which enters the current average can be written as
\begin{align}
G_{01}^K &= i \lambda \left(\sigma_0 - i \lambda^2 G_{00}^r \right) g_{00}^K \left( \sigma_0 + i\lambda^2 G_{00}^a  \right) \sigma_3 \nonumber  \\
&-2 i \lambda G_{00}^r F_1 \left(\sigma_0 + i \lambda^2  G_{00}^a \right) \sigma_3 ~.
\label{GO1K}
\end{align}
Where $G_{00}^{r/a}$ is also obtained from the Dyson's equation, and reads
\begin{equation}
G_{00}^{r/a} = \left[-\sigma_3 g_{00}^{r/a}\sigma_3 \pm i \lambda^2 \sigma_0\right]^{-1}~.
\label{G00RA}
\end{equation}
One readily sees that $G_{00}^{r/a}$ is a linear combination of $\sigma_0$ and $\sigma_1$ Pauli matrices
\begin{equation}
G_{00}^{r/a}(\omega) = \alpha^{r/a}(\omega) \sigma_0 + \gamma^{r/a}(\omega) \sigma_1 ~,
\end{equation}
where, denoting $\omega_{\pm} =  \left(\omega \pm i0^+\right)$,
\begin{equation}
\begin{array}{rl}
\alpha^{r/a}(\omega)  & = \dfrac{ \left( \sqrt{\Delta^2 -\omega_{\pm}^2} \mp i\lambda^2 \omega_{\pm}\right) \omega_{\pm}}{\Delta^2-\left(\sqrt{\Delta^2 - \omega_{\pm}^2}  \mp i\lambda^2 \omega_{\pm}\right)^2}~,  \label{alpha}
\end{array}
\end{equation}
\begin{equation}
\begin{array}{rl}
\gamma^{r/a}(\omega) & = \dfrac{\Delta \omega_{\pm} }{\Delta^2- \left(\sqrt{\Delta^2 - \omega_{\pm}^2}  \mp i\lambda^2  \omega_{\pm}\right)^2}~.
\label{gamma}
\end{array}
\end{equation}
Note that $\alpha^r = (\alpha^a)^*$ and $\gamma^r = (\gamma^a)^*$. 

Keeping in mind that $g_{00}^K$ is also a linear combination of $\sigma_0$ and $\sigma_1$, one can significantly simplify the Nambu trace needed for the current
\begin{align}
\mathrm{Tr_N} G_{01}^{K}(\omega) 
&= -2 i \lambda \mathrm{Tr_N} \left[ G_{00}^r F_1 \left(\sigma_0 + i \lambda^2  G_{00}^a \right) \sigma_3 \right] ~.
\end{align}
Inserting all these expressions in the trace formula for the current Eq. \eqref{tracecourant}, the result is, introducing $n_F(\omega\pm eV) \equiv n_F^{(\pm)}$ and $\delta n_F=n_F^{(+)} - n_F^{(-)}$:
 \begin{equation}
I(V) = - 2 i e \lambda^2 \int_{-\infty}^{+\infty} \frac{d \omega}{2\pi}  \left[\alpha^r + i\lambda^2 \left( |\alpha^r|^2 - |\gamma^r|^2\right)\right]\delta n_F~.
\label{current}
\end{equation}
This is identified as a Landauer formula
\begin{equation}
I = \dfrac{e}{2 \pi} \int_{-\infty}^{+\infty} \mathrm{d}\omega ~ \mathrm{R_A}(\omega) \left(n_F^{(-)} - n_F^{(+)}\right)~.
\label{eq:IvsRA}
\end{equation}
with the (real valued) Andreev reflection probability
\begin{equation}
\mathrm{R_A}(\omega) = - 2 \lambda^2 \left[ \text{Im} \left(\alpha^r\right) +  \lambda^2 \left( |\alpha^r|^2 - |\gamma^r|^2\right)\right]~,
\end{equation}
whose expression matches exactly with the one of the spectral current density $J(\omega)$ obtained in Eq.~(35) of  Ref. \onlinecite{zazunov_16}.

\section{Finite frequency noise}
\label{sec:noise}

\subsection{Real-time noise correlator}

The real time noise correlator in a multi-lead geometry is defined as:
\begin{equation}
S_{jl}(t,t') = \left\langle I_j(t) I_l(t') \right\rangle - \big\langle I_j(t) \big\rangle\left\langle I_l(t')\right\rangle~,
\label{noise}
\end{equation}
where $I_j(t)$ is defined by  Eq.~\eqref{currentop}. As the Hamiltonian is quadratic in the fermionic degrees of freedom, we can use Wick's theorem to express the four fermion operator averages in terms of products of off-diagonal Keldysh Green's functions, again in the form of a Nambu trace:    
\begin{align}
S_{jl}(t,t') &= e^2  \mathrm{Tr_N}\Big[  \sigma_3 \left( \hat{W} \hat{G}^{-+}(t,t') \hat{W}\right)_{jl} \sigma_3~ G^{+-}_{lj}(t',t) \Big]\nonumber\\
-  e^2 & \mathrm{Tr_N}\Big[ \sigma_3 \left( \hat{W} \hat{G}^{-+}(t,t')\right)_{jl} \sigma_3 \left( \hat{W} \hat{G}^{+-}(t',t)\right)_{lj} \Big]~.
\label{noisetrace}
\end{align}
where we recall that $\hat{\cdot}$ corresponds to matrices in Nambu-lead space.

In our two-terminal geometry, this simplifies as
\begin{align}
S_{11}(t,t') =  \lambda^2 e^2 \mathrm{Tr_N} & \left[ G^{-+}_{00}(t,t') G^{+-}_{11}(t',t) \right. \nonumber\\
  & \left. - G^{-+}_{01}(t,t') G^{+-}_{01}(t',t) \right]~.
\end{align}

These Green's functions are related to the retarded, advanced and Keldysh components according to 
\begin{align}
\hat{G}^{+-} & = \frac{1}{2} \left( \hat{G}^{a} - \hat{G}^r + \hat{G}^K \right) \label{G+-}~,\\
\hat{G}^{-+} & = \frac{1}{2} \left( \hat{G}^{r} - \hat{G}^a + \hat{G}^K \right) \label{G-+}~.
\end{align}
Writing explicitly the Keldysh component of the Dyson's equation, Eq.\eqref{eq:GK}, for all combination of leads, one has
\begin{align}
G_{00}^K & = \left( \sigma_0 - i  \lambda^2  G_{00}^r \right) g_{00}^K \left(  \sigma_0 + i \lambda^2  G_{00}^a \right) - 2 i \lambda^2  G_{00}^r F_1  G_{00}^a \\
G_{11}^K  &=   (\sigma_3 - i \lambda^2 \sigma_3 G_{00}^r)  (\lambda^2 g_{00}^K - 2 i F_1) (\sigma_3 + i \lambda^2 G_{00}^a \sigma_3)
\end{align}
The four Green's functions   $G^{-+}_{00}$, $G^{-+}_{01}$, $ G_{01}^{+-} $, $ G_{11}^{+-}$ which enter the real-time noise correlator are ultimately obtained from the above Keldysh components 
\begin{align}
G^{-+}_{00}    =&  \dfrac{1}{2} \Big[ G_{00}^r - G_{00}^a - 2 i \lambda^2 G_{00}^r F_1 G^a_{00} + H\Big]~,\\
G_{11}^{+-}  =  &i (\sigma_0 - F_1) + \dfrac{\lambda^2}{2}\sigma_3 \left(G_{00}^r - G_{00}^a\right) \sigma_3  \nonumber\\
  &- \lambda^2 \sigma_3 G_{00}^r\sigma_3 F_1+ \lambda^2 F_1 \sigma_3 G_{00}^a \sigma_3 \nonumber\\
  &- i\lambda^4 \sigma_3 G_{00}^r F_1 G_{00}^a \sigma_3+\frac{1}{2}\lambda^2\sigma_3 H \sigma_3~,\\
G^{-+}_{01}   = & - i\frac{\lambda}{2}  \left(G_{00}^r + G_{00}^a\right) \sigma_3 -  i\lambda G_{00}^r \sigma_3 F_1  \nonumber\\
 &+ \lambda^3 G_{00}^r F_1 G^a_{00} \sigma_3 + i\frac{\lambda}{2} H \sigma_3 ~,\\
G_{01}^{+-}  = &    i\frac{\lambda}{2} \left(G_{00}^r + G_{00}^a\right) \sigma_3 - i\lambda G_{00}^r \sigma_3 F_1 \nonumber\\
 &+ \lambda^3 G_{00}^r F_1 G_{00}^a \sigma_3 + i\frac{\lambda}{2} H \sigma_3 ~,
 \end{align}
where we introduced the new matrix
\begin{align}
H(\omega) =& \left( \sigma_0 - i \lambda^2 G_{00}^r  \right) g_{00}^K \left( \sigma_0 + i \lambda^2 G_{00}^a \right)~.
\end{align}
Substituting these  back into the expression for the real-time current correlator allows one to derive an exact, albeit cumbersome, analytic expression for $S_{11}$.


\subsection{Emission, absorption and measurable noise}

In full generality, one may consider that the Green's functions depend on two times $t$ and $t'$. In the specific setup under consideration, we are in a stationary situation as there is only one superconductor in the system. It follows that the time dependence reduces to a single variable, more precisely the time difference $\tau = t-t'$. We thus consider the two distinct correlators:
\begin{align}
S^+ (\tau) = S_{11} (0 , \tau) = \lambda^2 e^2 \mathrm{Tr_N}\Big[ & G^{-+}_{00}(-\tau) G^{+-}_{11}(\tau) \nonumber\\
&- G^{-+}_{01}(-\tau) G^{+-}_{01}(\tau)\Big]~, \\
S^- (\tau) = S_{11} (\tau , 0) = \lambda^2 e^2  \mathrm{Tr_N}\Big[ & G^{-+}_{00}(\tau) G^{+-}_{11}(-\tau) \nonumber\\
& - G^{-+}_{01}(\tau) G^{+-}_{01}(-\tau)\Big]~,
\end{align} 
Their Fourier transforms define the emission and absorption noise as
\begin{equation}
S^+ (\Omega) = \displaystyle\int_{-\infty}^{+\infty}\mathrm{d}t ~\langle \delta I(0) \delta I(t)\rangle ~\mathrm{e}^{i\Omega t}~, 
\end{equation}
\begin{equation}
S^- (\Omega) = \displaystyle\int_{-\infty}^{+\infty}\mathrm{d}t ~\langle \delta I(t) \delta I(0)\rangle ~\mathrm{e}^{i\Omega t}~. 
\end{equation}
which are conveniently rewritten using the Fourier-transformed Green's functions as
\begin{align}
S^+(\Omega)  = \lambda^2 e^2 \int_{-\infty}^{+\infty} \dfrac{\mathrm{d}\omega}{2\pi} & \mathrm{Tr_N} \left[ 
G^{-+}_{00}(\omega) G^{+-}_{11}(\omega+\Omega) \right. \nonumber\\
 & \qquad \left. -  G^{-+}_{01}(\omega) G^{+-}_{01}(\omega+\Omega) \right]
\label{Splus} \\
S^-(\Omega)  = \lambda^2 e^2 \int_{-\infty}^{+\infty} \dfrac{\mathrm{d}\omega}{2\pi} & \mathrm{Tr_N} \left[ 
G^{-+}_{00}(\omega) G^{+-}_{11}(\omega-\Omega) \right. \nonumber\\
 & \qquad \left. -  G^{-+}_{01}(\omega) G^{+-}_{01}(\omega-\Omega) \right]~.
\label{Smin}
\end{align} 
It is clear from these expressions that emission and absorption noises are trivially related when flipping the sign of the probing frequency, namely $S^+(\Omega) = S^-(-\Omega)$. It follows that in order to describe the whole range of physical parameters, we can safely focus on positive frequencies for both noise spectra.

In order to connect with potential experimental works looking to investigate the finite-frequency noise of the NTS junction, we also compute the noise expected to be measured from a generic quantum detector consisting of a resonant LC circuit, inductively coupled to the NTS junction (see Fig.~\ref{fig:system}). This measured noise is the result of repeated measurements of the charge at the capacitor plates. For the present setup, we use an expression formulated in Refs. \onlinecite{lesovik_97,zazunov_07} to predict the result of such a measurement. This so-called measurable noise is given by
\begin{equation}
S_{\mathrm{meas}}(\Omega) = K \{ S^+(\Omega) + N(\Omega) [S^+(\Omega) - S^-(\Omega)] \}~,
\label{Noisemeasure}
\end{equation}
where $N(\Omega)=[\exp(\hbar \Omega / k_B T_{LC}) - 1]^{-1}$ is the Bose-Einstein distribution associated with the oscillator modes, and K is the effective coupling constant of the quantum wire with the resonator whose expression reads
\begin{equation}
K = \Big(\dfrac{\alpha}{2 L}\Big)^2 \dfrac{1}{2\eta}~.
\end{equation}
$L$ and $\alpha$ respectively stand for the inductance of the resonant circuit and the inductive coupling, while $\eta$ is the width of the resonance. This last parameter depends on the dissipative environment surrounding the LC circuit,\cite{zazunov_07} and is typically small but its inverse is kept finite. Note that in full generality, this measurable noise is a combination of emission and absorption noise, weighted by the Bose- Einstein distribution. The temperature of the noise detector $T_{LC}$ can in general be different from that of the electrons flowing in the NTS junction. 

\section{Results}
\label{sec:numerical}

In this section, we start by briefly recalling the known results for the conductance and the zero frequency noise, as both these quantities serve as a useful benchmark for our approach. We then move on to the determination of the finite frequency emission, absorption and measurable noises at low temperature.

The NTS junction is characterized by its transparency $\tau$ ($0\leq\tau\leq1$), which is defined from the tunneling amplitude $\lambda$ as $\tau = 4 \lambda^2 / (1+\lambda^2)^2$. We use units such that  $k_B=\hbar=1$. We adopt the convention that the TS wire is grounded, the voltage drive being applied to the normal metallic lead. 
 
\subsection{Differential conductance}

We hereby recover important results which were already obtained in a previous study.\cite{zazunov_16}  
At zero temperature, the differential conductance is directly given by the reflection probability $G = \frac{2 e^2}{h} R_A (eV)$ [see Eq.~\eqref{eq:IvsRA}]. At subgap voltages, this takes a very simple form since in this case
\begin{align}
R_A (eV) = \frac{1}{1+ \left( \frac{eV}{\Gamma} \right)^2 }~, 
\label{eq:NTScond}
\end{align}
leading to the predicted Majorana zero-bias peak with quantized height $2e^2/h$ and width $\Gamma = \tau \Delta / 2 \sqrt{1-\tau}$ due to resonant Andreev reflexion. As one approaches perfect transmission $\tau \rightarrow 1$, the strong NTS hybridization turns this peak into a plateau at $G=2e^2/h$ extending over the whole subgap regime, $|eV|<\Delta$ (although the Majorana state is ill-defined due to this very same hybridization). Moreover, setting  $\Delta = 0$, one recovers the behavior of a spin polarized  N-N junction with a Landauer conductance $G_{NN} = \dfrac{\tau e^2}{h}$. Adding a finite temperature to the system does not substantially change the results and only tends to smooth out the curves, as expected.
	
\subsection{Zero-frequency noise}

We now turn to the shot noise $S_{11} = S^\pm (\Omega = 0)$, and more specifically the Fano factor, defined as the ratio of the zero-frequency noise to the average current, $F = S_{11} / eI_1$. 

In the subgap regime, as predicted in Ref. \onlinecite{zazunov_16}, this Fano factor vanishes at perfect transparency ($\tau=1$), and increases when the transparency decreases following
\begin{align}
F = 1 - \frac{1}{\arctan \left( \frac{eV}{\Gamma} \right)} \frac{\frac{eV}{\Gamma}}{1 + \left( \frac{eV}{\Gamma} \right)^2}~.
\end{align}

Above the gap, even at $\tau=1$, the Fano factor $F$ is finite due to the concomitant presence of both Andreev reflexion and quasiparticle processes. 

For a normal-normal junction, i.e. in the limit $\Delta=0$, the zero-frequency noise at zero temperature shows, as expected,\cite{blanter_00,martin_05} a linear behavior with respect to the bias voltage and reduces to  $S_{11} = (e^2/h) \tau (1-\tau) eV$. 

\subsection{Finite-frequency noise}

The finite-frequency noise has so far been absent in the investigations of the transport characteristics of topological superconductors. The results for the emission noise $S^+(\Omega)$ and the absorption noise $S^-(\Omega)$, obtained from a numerical evaluation of Eqs.~\eqref{Splus} and \eqref{Smin}, are shown in this section. Only their real parts are discussed here, as their imaginary parts yield vanishing contributions within numerical accuracy.
 
\begin {figure}[tb]
\centering
\includegraphics[width=\columnwidth]{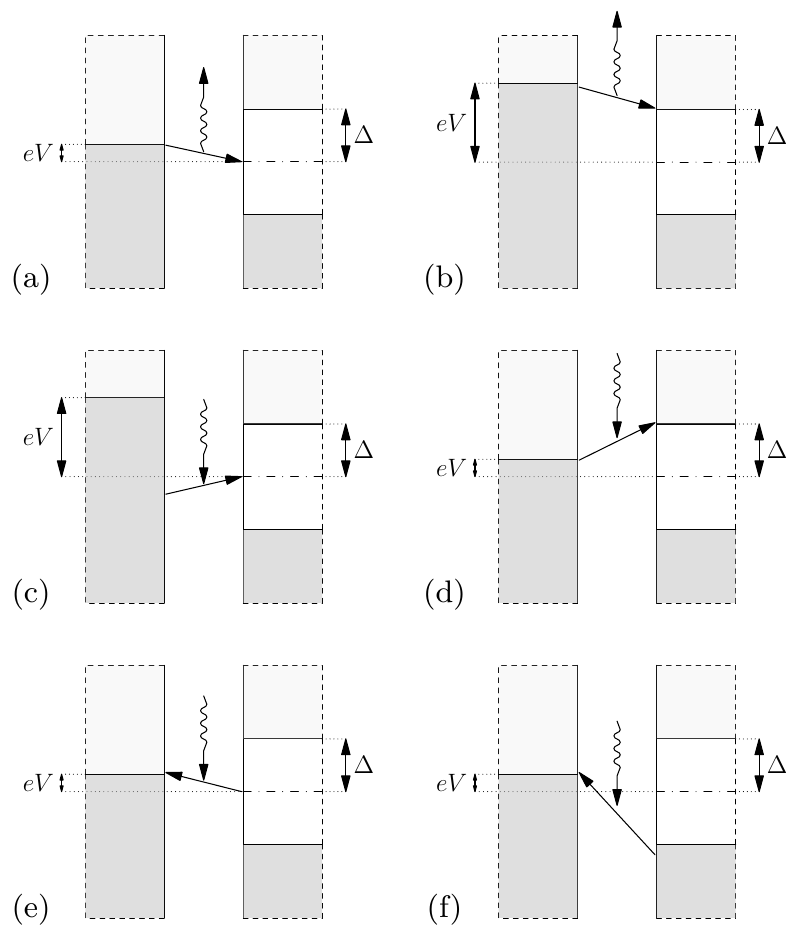} 
      \caption{Energy diagram of a NTS junction. In medium grey are represented the occupied electronic states and in light grey the empty states. (a)-(b) emission processes, (b) is not available for a bias $eV$ below the gap $\Delta$, (c)-(d) absorption processes involving a transition from the metal to the superconductor, (e)-(f) absorption processes with transitions to the metal. Straight lines correspond to the transfer of electrons while wiggly lines are associated with the absorption or emission of photons.}
      \label{SchemaNTS}
   \end {figure} 
 
  The noise at finite frequency $\Omega$ is created by fluctuations
of current processes which are accompanied by the emission or absorption of a photon at
the frequency $\Omega$. Understanding the basic process which contributes to the current
with emission or absorption of a photon, while conserving the total energy, can give us a qualitative understanding of the finite-frequency noise features. 
   
   As a warm-up, let us first recall what happens in the most simple situation of a junction between two normal metallic electrodes (corresponding to taking the limit $\Delta \to 0$). There, electronic transport occurs if occupied electron states from one lead have enough energy to reach the empty states of the other. This results in a decreasing linear behavior of the emission noise $S^+(\Omega)$ for frequencies $\Omega \in~ [0,eV]$  and a vanishing emission noise for higher frequencies due to Pauli blocking.\cite{martin_05} In the meantime, the absorption noise  $S^-(\Omega)$ increases linearly with $\Omega$, since the bigger the frequency, the more electrons can be transfered between leads.
   
When we replace the normal lead by a topological superconducting wire, the noise spectrum changes drastically. These changes can be understood qualitatively by taking into account the density of states of the topological superconductor, which
is gapped for energies $|E|<\Delta$, except for a single narrow peak at zero energy due to the Majorana bound state
(see Fig.~\ref{fig:system}). 
 
\subsubsection{Qualitative picture}

The qualitative picture showing the basic processes contributing to the finite-frequency
noise are shown in Fig.\ref{SchemaNTS}. The two processes of the top row (a and b)
describe emission noise, while the four other processes  describe absorption noise.
Note that we can choose $eV>0$ without loss of generality, thanks to the electron-hole symmetry.

\begin{figure}[tb]
\centering
\includegraphics[width=\columnwidth]{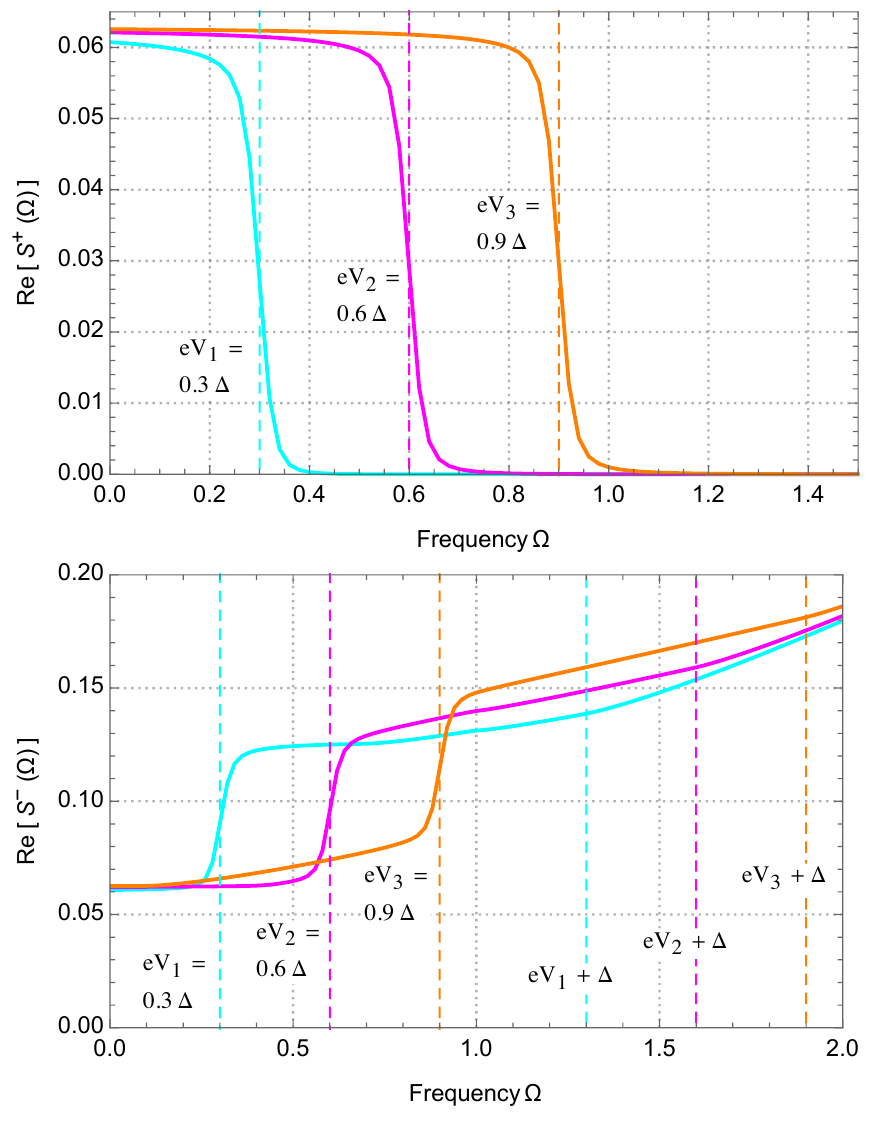}
\caption{Emission noise $S^+(\Omega)$ (top) and absorption noise $S^-(\Omega)$ (bottom) as a function of frequency $\Omega$ (in units of $\Delta$) at low transparency ($\tau=0.02$) computed in the subgap regime $|eV|<\Delta$ for three different values of the voltage bias ($eV = 0.3 \Delta$, $0.6 \Delta$ and $0.9 \Delta$), and expressed in units of $e^2 \Delta/h$.}
	   \label{S002subgap}
\end{figure}

When the bias voltage is smaller than the gap, $eV<\Delta$,
there is only one process contributing to the emission noise, which is shown on panel (a). Electrons in the normal
metal with energies in the interval $[0,eV]$ may hop to the Majorana state by emitting a photon of energy $\Omega = eV$ (see Fig.~\ref{SchemaNTS}a).  Since the width of the Majorana peak in the density of states is extremely small, and the DOS of the normal metal is constant, this emission process only shows a negligible dependence on the frequency.

For voltages larger than the gap, another emission process kicks in:
electrons from the normal metal can be transferred to the empty states above the TS gap,
and emit a photon with $\Omega$ in the range $[0,eV-\Delta]$ (see Fig.~\ref{SchemaNTS}b).
The frequency dependence of this process will reflect the energy dependence of the TS density
of states above the gap.

Absorption processes similar to the emission processes discussed above can also be realized. These are shown
on panels (c) and (d) of Fig.\ref{SchemaNTS}, and are expected to show similar frequency dependence as their emission counterparts. In the case of panel (c), an electron from the normal metal absorbs a photon to be transmitted to the Majorana state. Panel (d) is the equivalent of panel (b) for photon absorption: an electron from the normal metal absorbs a photon and is transmitted to an empty state above the TS gap. Note however that while processes (a) and (c) exist for any voltages, the emission process of panel (b) can only occur for voltages beyond the gap.

Finally, the absorption of photons also allows processes where an electron is transmitted from the topological
superconductor to the normal metal, for frequencies $\Omega \in~[eV,+\infty]$. Panels (e) and (f) of Fig.\ref{SchemaNTS} show the
corresponding processes, where the TS electron is transmitted to a normal metal empty state, either from the Majorana state [panel (e)] or from the occupied states below the gap [panel (f)].

\subsubsection{Results for $eV < \Delta$}
	
We now show the results for the finite frequency emission noise $S^+(\Omega)$ and absorption noise $S^-(\Omega)$ obtained at low temperature $T_e=0.01\Delta$ for different values of the transparency $\tau$ and the voltage $V$ (applied on the normal lead, the TS wire is assumed to be grounded).

The emission noise in the subgap regime $eV<\Delta$ is shown in Fig.~\ref{S002subgap}~(top), at low transparency $\tau=0.02$ for three different values of the voltage. 
For each curve, we observe a plateau that extends up to $\Omega= eV$, followed by a sharp drop to zero. This plateau corresponds to the process (a) of Fig.\ref{SchemaNTS}, and exists only for frequencies $\Omega < eV$, hence the observed drop. The absorption noise for the same parameters is shown on Fig.~\ref{S002subgap}~(bottom). The most visible feature is a sharp
step at $\Omega = e V$, which corresponds to the onset of the process (e) from
Fig.~\ref{SchemaNTS}, whose contribution is independent of frequency. 

Smaller structures can be observed at frequency $\Omega= \Delta - e V$ and $\Omega= \Delta + e V$, corresponding respectively to the onset of processes (d) and (f) from Fig.~\ref{SchemaNTS}. Interestingly, a fourth process also contributes to the absorption noise, namely process (c). However, the latter cannot be associated with any visible feature as it leads to a constant contribution for all positive frequency $\Omega$, thus resulting in a finite background contribution to the absorption noise.

\begin{figure}[tb]
\centering
\includegraphics[width=\columnwidth]{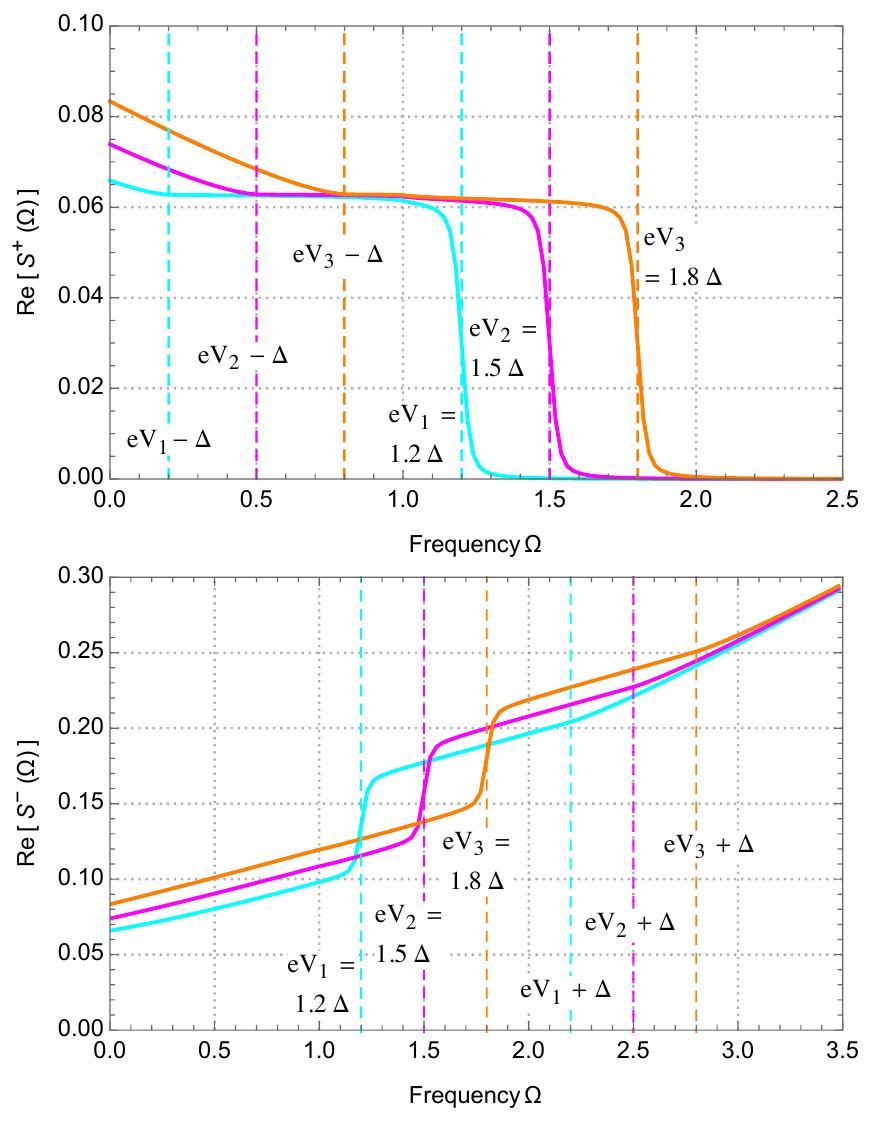}
\caption{Emission noise $S^+(\Omega)$ (top) and absorption noise $S^-(\Omega)$ (bottom) as a function of frequency $\Omega$ (in units of $\Delta$) at low transparency ($\tau=0.02$) computed beyond the gap $|eV|>\Delta$ for three different values of the voltage bias ($eV = 1.2 \Delta$, $1.5 \Delta$ and $1.8 \Delta$), and expressed in units of $e^2 \Delta/h$.}
	   \label{S002abovegap}
\end{figure}

\subsubsection{Results for $eV > \Delta$}

We now turn to the low-temperature results for the noise at voltages $eV>\Delta$ and in the same low transparency regime ($\tau=0.02$). 

The emission noise $S^+ (\Omega)$ is shown in Fig.~\ref{S002abovegap}~(top), for three different voltages beyond the gap.  As in the subgap regime, the main feature of the emission noise is the presence of a well-defined plateau extending from $\Omega = eV - \Delta$ all the way up to $\Omega = eV$, which ends with a sharp drop to zero. Again, this plateau is fully compatible with the process shown in Fig.~\ref{SchemaNTS}a, as it is expected to lead to a constant contribution, and can only occur for frequencies $\Omega < eV$ (since there are no occupied electronic states at higher energy). While this contribution is still present at very low frequency, it is supplemented by another one arising from the process (b) (see Fig.~\ref{SchemaNTS}) leading to a new structure in the noise. Indeed, for frequencies $\Omega < eV - \Delta$, electrons close to the metallic Fermi level can be transfered to the empty states above the TS gap while emitting a low-energy photon. The resulting frequency dependence of the emission noise is directly related to the energy dependence of the TS density of states above the gap. 

The absorption noise for $e V>\Delta$ is shown on Fig.~\ref{S002abovegap}~(bottom). 
The main feature
is again a sharp step for $\Omega= eV$, associated with the process presented in panel (e) of Fig.~\ref{SchemaNTS}. A smaller structure is visible at $\Omega= \Delta + e V$, and can be directly tied to the onset of process (f). Processes (c) and (d) also contribute to the absorption noise, being present at all frequencies and leading respectively to a constant background contribution and to a steady increase of the noise.

\begin{figure}[tb]
   \centering
   \includegraphics[width=\columnwidth]{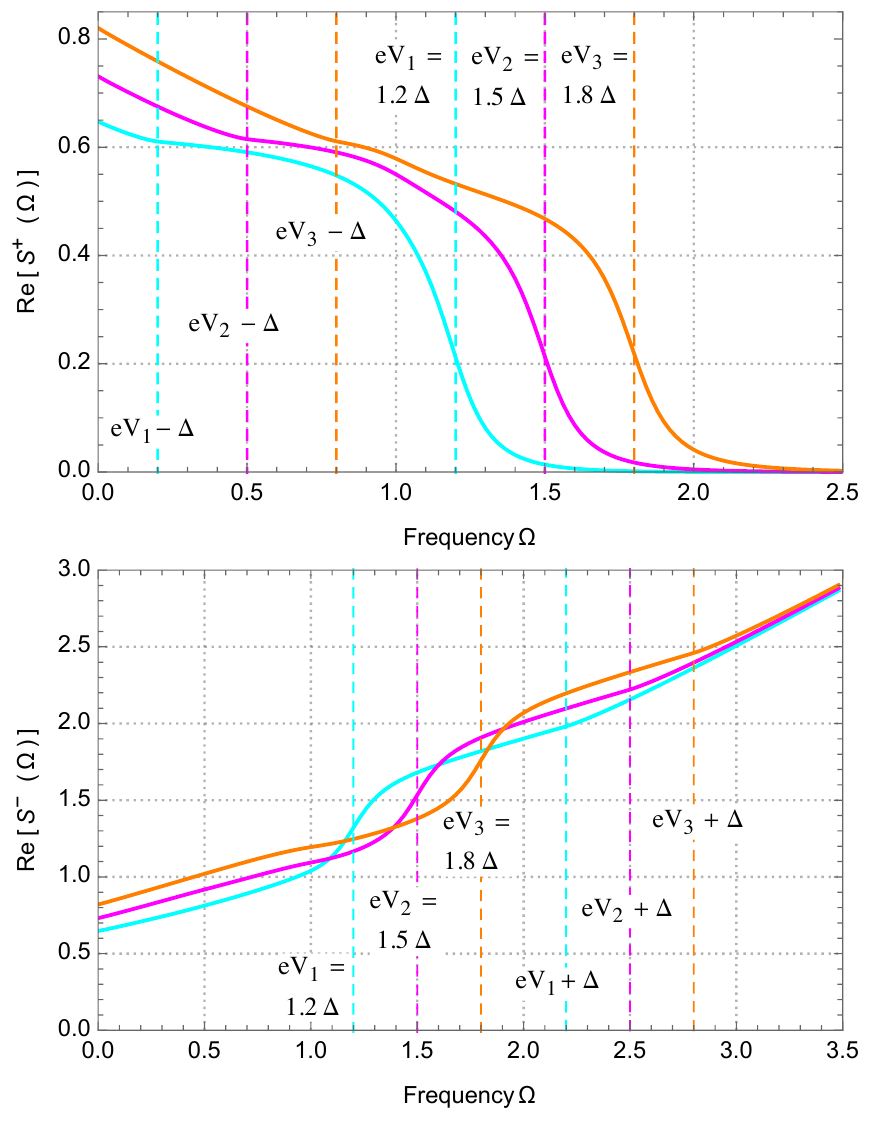}
\caption{Emission noise $S^+(\Omega)$ and absorption noise $S^-(\Omega)$ (in units of $e^2 \Delta/h$) as a function of frequency $\Omega$ (in units of $\Delta$) at transparency $\tau=0.4$ beyond the gap $|eV|>\Delta$, for three different values of the voltage bias ($eV = 1.2 \Delta$, $1.5 \Delta$ and $1.8 \Delta$).}
	   \label{S04}
\end{figure}

It follows from the inspection of the finite-frequency noise, for voltages both above and below the TS gap, that the presence of a Majorana bound state leads to a plateau in the emission noise as a function of frequency, accompanied by a sharp drop to zero at $\Omega = eV$.

At this point, it is interesting to relax the constraint of low transparency and study the impact of a higher value of $\tau$. This is shown in Fig.~\ref{S04}, where an intermediate value of the transparency, $\tau =0.4$, has been used. 

One can readily see that increasing the transparency tends to smooth out all the structures visible in the noise and discussed above. Further increasing the transparency and ultimately approaching $\tau=1$ (not shown) makes all features in the noise spectrum disappear. The best opportunity to identify the noise plateau associated with the Majorana bound state is thus to focus on the case of tunnel junctions.

\subsection{Measurable noise}
	
We end this study with a short discussion of the prediction of the measurable noise. To this end, we choose the inductive coupling scheme introduced in Ref.~\onlinecite{lesovik_97} which involves an LC circuit in the vicinity of the NTS junction (see Fig.~\ref{fig:system}). This protocol relies on repeated measurements of the charge on the plates of the capacitor of the LC resonant circuit. We aim at investigating the results of the measurable noise $S_\text{meas} (\Omega)$ as a function of the frequency $\Omega$, as defined in Eq.~\eqref{Noisemeasure}, which corresponds to the measurable response of this resonant circuit. 

\begin{figure}[tb]
\centering
\includegraphics[width=\columnwidth]{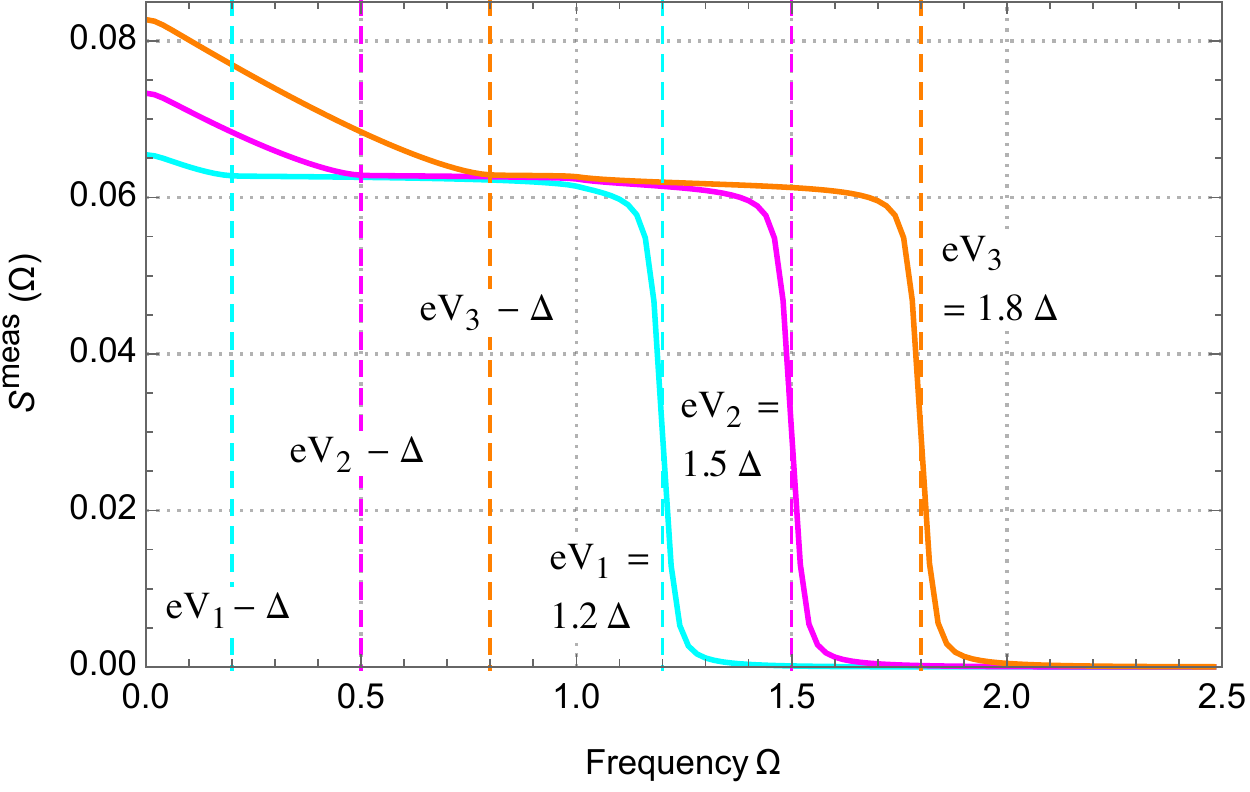}
\caption{Measurable noise (in units of $e^2 K \Delta/h$) as a function of frequency $\Omega$ (in units of $\Delta$) at low transparency ($\tau=0.02$) for electron and detector temperatures $T_e = T_{LC}=0.01\Delta$.}
	   \label{Smeaslowtemp}
\end{figure}

Following our investigations on the emission and absorption noise, we focus here on the low transparency regime, as this is the most likely candidate to observe meaningful signatures of the Majorana bound state. The new crucial parameter which dramatically influences the measurable noise is the temperature $T_{LC}$ of the detector.

In the low temperature regime for the detector, $T_{LC} \ll \Omega$, the measurable noise, presented in Fig.~\ref{Smeaslowtemp} for voltages beyond the gap, closely resembles the emission noise obtained earlier for the same set of parameters, showing in particular the characteristic low-frequency behavior of the noise which we could associate with the presence of a Majorana bound state. This is easily understood from the very definition of the measurable noise, Eq.~\eqref{Noisemeasure}. Indeed, in this low temperature regime, the Bose-Einstein distribution becomes vanishingly small, so that the measurable noise reduces to
\begin{align}
S_\text{meas} \left(\Omega \gg T_{LC} \right) \simeq K S^+ (\Omega)~.
\end{align}

\begin{figure}[tb]
\centering
\includegraphics[width=\columnwidth]{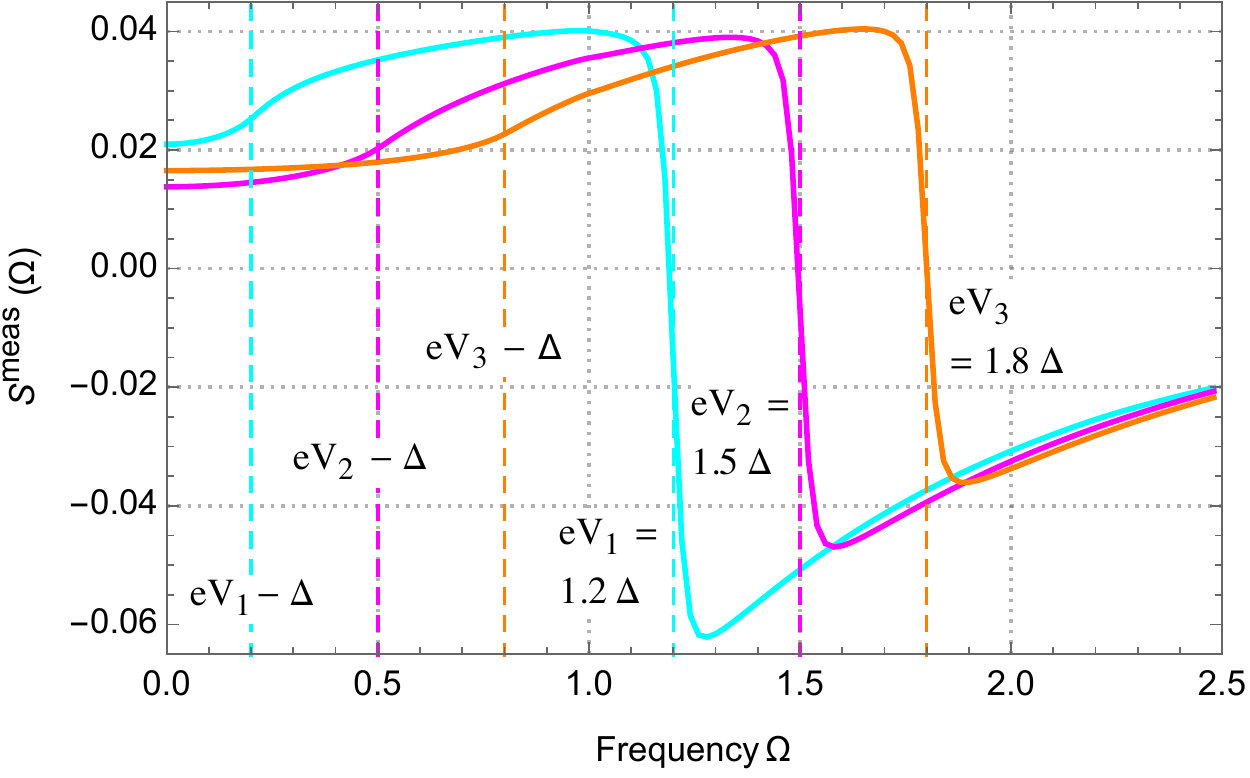}
\caption{Measurable noise (in units of $e^2 K \Delta/h$) as a function of frequency $\Omega$ (in units of $\Delta$) at low transparency ($\tau=0.02$) for an electron temperature $T_e =0.01\Delta$ and a detector temperature $T_{LC}=\Delta$.}
	   \label{Smeashightemp}
\end{figure}

As one increases the detector temperature (see Fig.~\ref{Smeashightemp} obtained for $T_{LC} = \Delta$), this connection with the emission noise is progressively altered, ultimately leading to the disappearance of any meaningful signatures. However, the typical onsets of the various processes stay unaffected leading to structures in the measurable noise at frequencies $\Omega = eV - \Delta$ and $\Omega = eV$. In particular,  the measurable noise shows a sudden change of sign near $\Omega = eV$, a feature inherited from the sharp drop in the emission noise, which is also directly connected to the presence of the Majorana peak in the TS DOS. It follows that even in this high temperature regime of the detector, the measurable noise can still be used as a tool to uncover signatures of the Majorana bound state.
	 
\section{Comparison with a N-dot-S system}
\label{sec:NdS}

A remaining question is to inquire whether a non-topological superconductor/normal metal junction bearing zero-energy Andreev bound states (ABS), which also exhibits a zero bias peak in the differential conductance, is likely to produce the same finite-frequency noise characteristics as the NTS junction, which constitutes the focus of this study. We thus consider here a system where a standard BCS superconductor is connected to a normal lead via a single quantum dot. We compute the finite frequency-noise in a regime
where the quantum dot has signatures close to a Majorana bound state when looking at the conductance only. This will allow us to point out the unique information obtained by considering the finite-frequency noise, which permits to make a clear distinction between a real Majorana bound state in a topological superconductor and the presence of an accidental bound state at zero energy close to the junction.

The Hamiltonian of the system is:
\begin{equation}
H = H_D + \sum_{j=S,N} H_j + H_T
\end{equation}
where $H_D$ is the Hamiltonian of the dot level at energy $\epsilon$:
\begin{equation}
H_D = \epsilon \sum_{\sigma=\uparrow,\downarrow} d^{\dagger}_{\sigma} d_{\sigma}   ,
\end{equation}
$H_j$ ($j=S,N$)  are the Hamiltonians of the BCS superconductor and of the normal lead:
\begin{equation}
H_j = \sum_k \Psi^{\dagger}_{j,k} \left(\xi_k \sigma_z + \Delta_j \sigma_x \right)
  \Psi_{j,k} 
\end{equation}
where $\Delta_S=\Delta$ is the superconducting gap, and $\Delta_N=0$. 
$\Psi_{j,k}$ is a Nambu spinor:
\begin{equation}
\Psi_{j,k}= \left( \psi_{j,k,\uparrow}  \quad  \psi^{\dagger}_{j,-k,\downarrow} \right)^T ,
\end{equation}
$\xi_k = k^2/(2m) - \mu$, and $\sigma_x$ and $\sigma_z$ are Pauli matrices in Nambu space.
Finally $H_T$ is the tunneling Hamiltonian between the leads and the dots.
\begin{equation}
H_T = \sum_{jk} \Psi^{\dagger}_{jk} \mathcal{T}_j(t) d + \mbox{H.c.} 
\end{equation}
where $d= (d_{\uparrow} \; d^{\dagger}_{\downarrow})^T$ is a Nambu spinor for the dot electrons. $\mathcal{T}_j$ is the tunneling amplitude between lead $j$ and the dot,
where the constant voltage $V_j$ has been included via a Peierls substitution :
\begin{equation}
\mathcal{T}_j(t) = t_j \sigma_z \mbox{exp}( i \sigma_z V_j t)
\end{equation}
The tunneling rate between the dot and lead $j$ is further defined as $\Gamma_j = \pi \nu_0 t_j^2$, where $\nu_0$ is the density of states of the metal at Fermi energy.

Calculation of the average current and current correlations is carried out using the Keldysh Green function formalism, and has already been presented for a similar system in Ref.~\onlinecite{chevallier2011}. The formula for the mean current to lead $j$ is:
\begin{equation}
\left \langle I_j \right \rangle = e \text{Tr}_N \left\{
  \sigma_3 \int  \frac{d\omega}{2\pi} \mbox{Re}
  \left[ \cG^{r}(\omega) \Sigma^{K}(\omega) + \cG^{K}(\omega) \Sigma^{a}(\omega) \right] 
\right\}
\label{eq:IjNdS}
\end{equation}
where the trace is performed in Nambu space, $\cG$ is the full dot Green function (including
the coupling to the leads), and $\Sigma_j$ is the self-energy for lead $j$. The superscripts
$r,a,K$ refer to the retarded, advanced and Keldysh components.

\begin{widetext}
Similarly, the formula for the current autocorrelations at finite frequency $\Omega$ is:
\begin{align}
S_{j}(\Omega) &= - \frac{e^2}{2} \text{Re} \int \frac{d \omega}{2 \pi} \text{Tr}_{N} \left\{ \sigma_3 \left[\Sigma_j^K \cG^a + \Sigma_j^r \cG^K - \Sigma_j^a \cG^a + \Sigma_j^r \cG^r \right] _\omega
\sigma_{3} \left[ \Sigma_j^K \cG^a + \Sigma_j^r \cG^K + \Sigma_j^a  \cG^a - \Sigma_j^r  \cG^r \right]_{\omega+ \Omega} \right.\nonumber\\
&\qquad \qquad \qquad \qquad \left. - \sigma_3 \left[ \Sigma_j^r \cG^r \Sigma_j^K + \Sigma_j^K \cG^a \Sigma_j^a + \Sigma_j^r \cG^K \Sigma_j^a
- \Sigma_j^a \cG^a \Sigma_j^a + \Sigma_j^r \cG^r \Sigma_j^r \right]_\omega
\sigma_3 \left[ \cG^K + \cG^a - \cG^r \right]_{\omega+ \Omega} \right\} \nonumber \\
&-  \frac{e^2}{4}  \text{Re} \int \frac{d \omega}{2 \pi} \text{Tr}_{N} \left\{ \sigma_3 \left[ \Sigma^a_j - \Sigma^r_j - \Sigma^K_j \right] _\omega  \sigma_3 \left[\cG^a - \cG^r + \cG^K \right]_{\omega+ \Omega}  +  \sigma_3 \left[ \cG^r - \cG^a + \cG^K \right]_\omega \sigma_3 \left[\Sigma^r_j -\Sigma^a_j  - \Sigma^K_j \right]_{\omega+ \Omega} \right\}.
\label{eq:SjNdS}
\end{align}
Note that there is an extra term compared to Eq.~(51) of Ref.~\onlinecite{chevallier2011}, as we are not computing the current cross-correlations between different leads, but instead the autocorrelations.
For $\Omega>0$ (resp. $\Omega<0$), Eq.~\eqref{eq:SjNdS} gives the emission noise $S^+(\Omega)$ [resp. the absorption noise $S^-(\Omega)$] - to be compared with Eqs.~\eqref{Splus}-\eqref{Smin} for the NTS case.
\end{widetext}

\begin{figure}
\centerline{\includegraphics[width=7.cm]{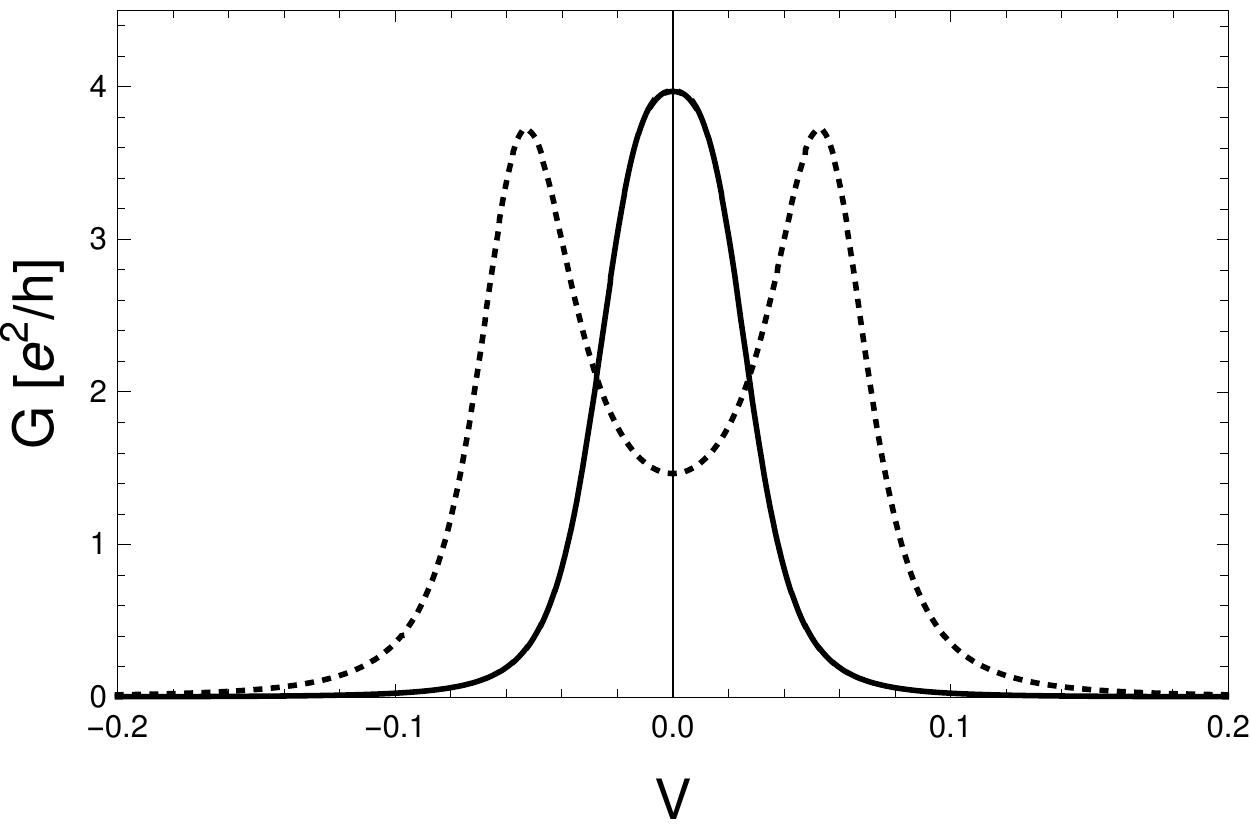}}
\caption{Differential conductance $G=dI_N/dV$, for the current to the normal lead $I_N$, in a
N-dot-S junction, with symmetric couplings ($\Gamma_N=\Gamma_S=0.02$, full line) and
asymmetric couplings ($\Gamma_N=0.02$, $\Gamma_S=3 \Gamma_N$, dotted line), as a function
of the voltage of the normal lead. Other parameters : $\epsilon=0$, $1/(k_B T)= 300$. 
The superconducting gap $\Delta$ is the unit of energy.}
\label{fig:INdS}
\end{figure}

We now turn to the results obtained with Eqs.~\eqref{eq:IjNdS} and \eqref{eq:SjNdS}. 
As a first step, we need to specify the parameters of the N-dot-S system for which the conductance is similar to the one
obtained for a NTS junction due to the presence of the Majorana bound state. This means
a conductance peak at zero energy, as shown by Eq.~\eqref{eq:NTScond}. 
This can be obtained here if two conditions are satisfied: setting the energy $\epsilon$ of the dot to zero, and using symmetrical couplings ($\Gamma_N = \Gamma_S$). While the first condition is obvious, the second one is illustrated
in Fig.~\ref{fig:INdS}, which shows the differential conductance G=$dI_N/dV$ as a function of the voltage $V$ applied to the normal lead. There, we consider the case of a symmetric junction with $\Gamma_N=\Gamma_S=0.02$ (full line), as well as the situation of
an asymmetric junction with $\Gamma_N = 0.02$ and $\Gamma_S= 3 \, \Gamma_N$ (dotted line). The other parameters
are  $\epsilon=0$, and $1/(k_B T) = 300$, while the superconducting gap $\Delta$ is the unit of energy. One can see that while the symmetric junction leads to a conductance peak which is identical to the one of a NTS junction, the asymmetric one shows a splitting of the conductance peak. 

\begin{figure}
\centerline{\includegraphics[width=8.cm]{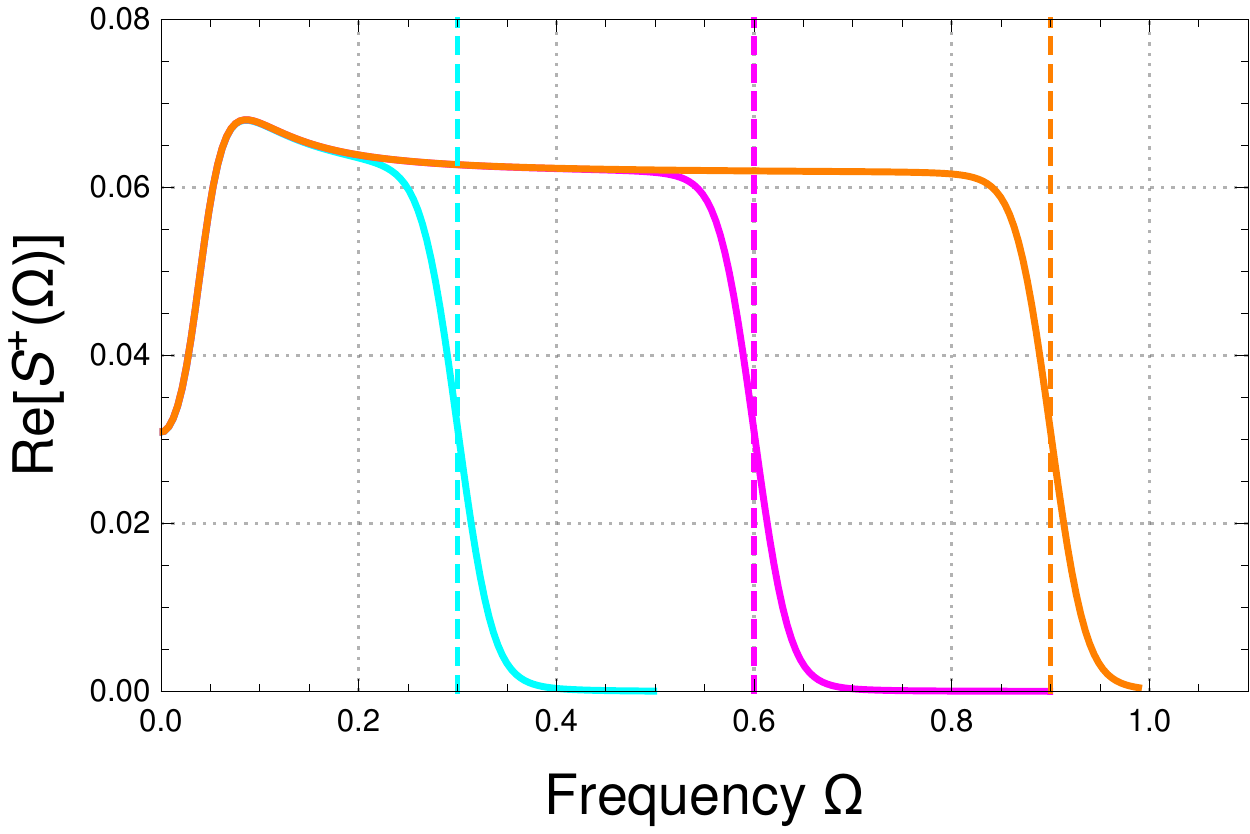}}
\caption{Emission noise $S^{+}(\Omega)$ for a symmetric N-dot-S junction with $\Gamma_N=0.02$,
 at voltages $V=0.3 \Delta, 0.6\Delta$ and $0.9\Delta$ (indicated by dashed vertical lines). Compare with Fig.~\ref{S002subgap} for the case of a NTS junction.}
\label{fig:SNdS1}
\end{figure}

\begin{figure}
\centerline{\includegraphics[width=8.cm]{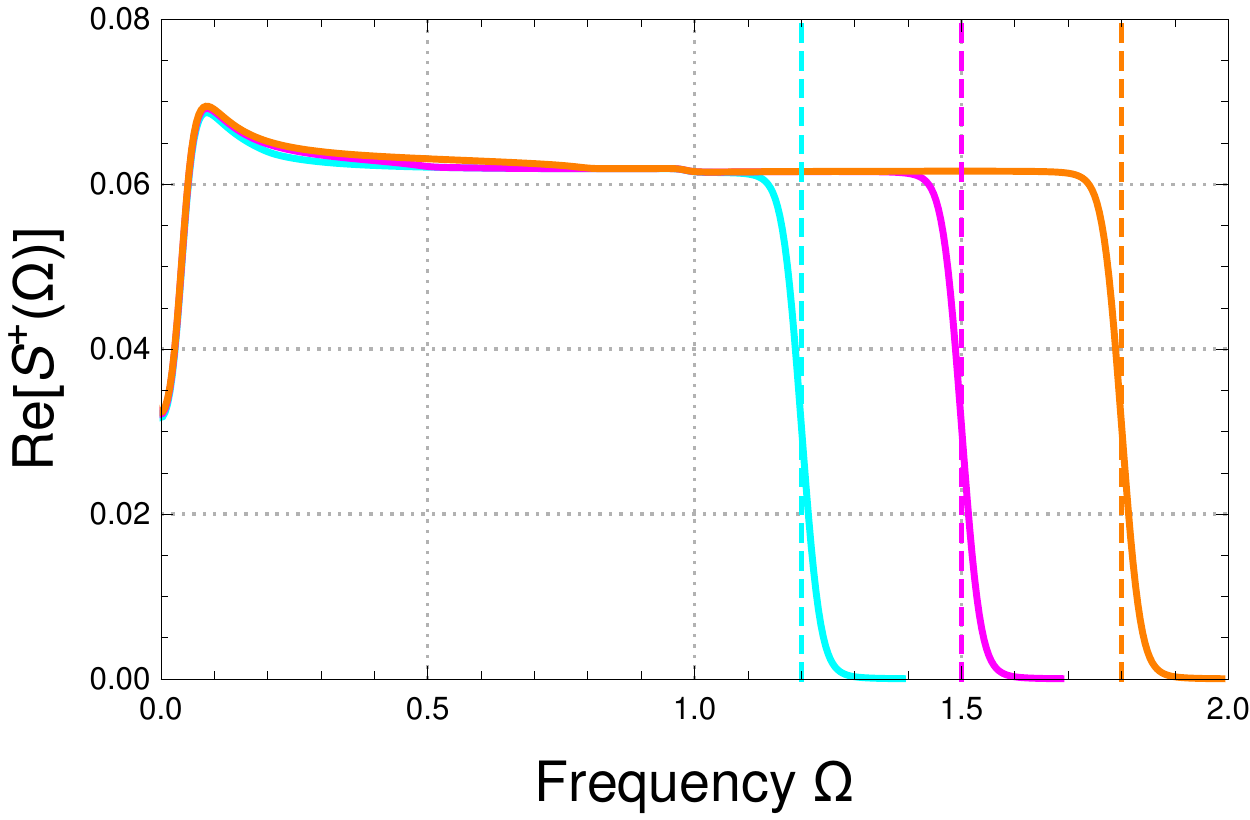}}
\caption{Emission noise $S^{+}(\Omega)$ for a symmetric N-dot-S junction with $\Gamma_N=0.02$,
 at voltages $V=1.2\Delta, 1.5\Delta$ and $1.8\Delta$ (indicated by dashed vertical lines). Compare with Fig.~\ref{S002abovegap} for the case of a NTS junction.}
\label{fig:SNdS2}
\end{figure}

Focusing then on a symmetric junction with a dot at zero energy ($\epsilon=0$), we now plot the emission noise.
Fig.~\ref{fig:SNdS1} shows $S^{+}(\Omega)$ for three values of the voltage below the gap
($V=0.3 \Delta, 0.6\Delta$ and $0.9\Delta$),
while Fig.~\ref{fig:SNdS2} focuses on voltages above the gap ($V=1.2\Delta, 1.5\Delta$ and $1.8\Delta$), with couplings $\Gamma_N=\Gamma_S=0.02$ in both cases.  These figures should be compared with their NTS equivalent, Fig.~\ref{S002subgap} and  Fig.~\ref{S002abovegap} respectively.

We see that, as in the case of the NTS junction, there is a clear plateau in the emission noise, which extends up to $\Omega= eV$. This plateau is the manifestation of the discrete level (here the dot level), which is associated with a narrow peak in the density of states.
However, an important difference with respect to the NTS case is also visible in Figs.~\ref{fig:SNdS1} and \ref{fig:SNdS2}:
at small frequency $\Omega$, there is a dip in the emission noise, with a value at $\Omega=0$ which is exactly half that of the plateau. 
  When probed on a very long timescale (as in the zero frequency noise), the current fluctuations are reduced by a factor $1/2$ due to the double-barrier structure with symmetric couplings. When the timescale is reduced to a time smaller than the average transfer time of an electron between an electrode and the dot level ($\sim 1/\Gamma$), then the double-barrier nature of the setup has no impact on the current fluctuations, which explains why the dip in $S^{+}(\Omega)$ has a width $\sim \Gamma$. This factor $1/2$ for the contrast of the observed dip corresponds to the well-known Fano factor reduction in double-barrier symmetric junctions.\cite{martin_05}
  Interestingly, this dip in the emission noise is completely absent in the case of a NTS junction, because the Majorana bound state is an intrinsic part of the superconductor, and is not separated by a barrier. It follows that, unlike the differential conductance, the low-frequency behavior of the emission noise is able to discriminate between the NTS and the N-dot-S system. Granted, our comparison between the noise characteristics of the NTS junction and this system containing a zero energy Andreev bound state is specific to the N-dot-S model we have adopted here. More generally, zero energy Andreev bound states can be generated in the contact region due to disorder and/or to multiple subbands in the nanowire. Studying this more general situation in detail goes well beyond the scope of the present paper. At this point, we cannot absolutely rule out that such an embedded ABS in the superconductor would lead to a frequency noise characteristics similar to that of the NTS junction. We therefore cannot claim that the frequency noise characteristics obtained in this work can uniquely be attributed to the presence of the Majorana fermion in the NTS junction. 


\section{Conclusion and perspectives}
\label{sec:conclusion}

	 The hunt for Majorana fermions continues to be an active field of research. Besides the zero voltage anomaly which was predicted, and in principle measured, by several experimental and theoretical groups, more signatures of these manifestations of collective electronic excitations are desperately needed. 
	 
In this work, we focused on a normal metal/topological superconductor junction subjected to a constant voltage bias using the non-equilibrium Keldysh formalism. Starting from the boundary Green's functions describing the two isolated parts, we were able to express the average current and noise in terms of dressed Green's functions, accounting for tunneling between the leads at all orders. This allows us to investigate the stationary current, the zero frequency noise and, the main result of this analysis, the finite-frequency emission and absorption noise. 

Computed for various voltages, at small and intermediate transparency, the finite-frequency noise $S^{+}(\Omega)$ and $S^{-}(\Omega)$ for the NTS junction show various structures, which can be tied to different emission or absorption processes. In particular, the distinctive feature associated with the presence of the Majorana bound state is a plateau in the noise, extending all the way from $\Omega=0$ up to a sharp step at frequency $\Omega =e V$ for voltages in the subgap regime. The presence of such a plateau is a general feature of a system with a discrete level, which is here the Majorana bound state. 
However, for a non-topological system with an accidental discrete level at zero-energy, we have shown by computing the emission noise in a N-dot-S system that the double barrier nature of the setup imposes a dip in the plateau at low frequency $\Omega$, which is absent for the NTS junction. In this particular N-dot-S system, the observation of the plateau for the whole frequency range could thus be used to discriminate between an accidental zero-energy Andreev bound state, and a real Majorana bound state. Nevertheless, this still leaves open the possibility that more general superconducting systems which bear ABS embedded in the superconductor would generate the same differential conductance and noise signals as the NTS junction. 
In any case, tunnel junctions should still be favored, as these characteristic structures are more prominent in the limit of small tunneling amplitudes, and get smoothed out when the transparency is increased.

Finally, we used a generic model for measuring noise consisting of a LC circuit inductively coupled to the normal metal/topological superconductor junction under study. In actual experiments, the temperature of this LC circuit is not necessarily the same as the electronic temperature of the junction. We see that when the temperature of the detection (LC) circuit is comparable to that of the electronic temperature of the NTS junction, we recover the same features as the ones observed in the emission and absorption noise. While higher temperatures of the detection circuit slightly blur the quality of this diagnosis, there still remains specific signatures which can be tied to tunneling processes involving the Majorana bound state. Our results therefore suggest that finite-frequency noise measurements offer a novel diagnostic of the presence of Majorana fermions in condensed matter systems. 
Extensions could include taking into account the finite length of the TS wire where the two Majorana bound states located at each end hybridize, or changing the chemical potential of the continuum version of the Kitaev chain model used here in order to trigger a transition to the non-topological phase. Using a more realistic model for the topological superconductor, including the electron spin, induced superconductivity, etc. is another notable extension in view of a quantitative comparison with experiments.\cite{zazunov17}

\begin{acknowledgments}
The project leading to this publication has received funding from Excellence Initiative of Aix-Marseille University - A*MIDEX, a French "Investissements d'Avenir" program.
\end{acknowledgments}

\end{document}